\documentstyle[aps,epsf]{revtex}

\begin{document}
\noindent{\footnotesize Lectures
presented at the International School on ``Pair Correlation in Many-Fermion-Systems'',\\
Erice (Sicily), June 4 -- 10, 1997; to be published in {\it Pair Correlation in Many-Fermion-Systems},\\
ed. V. Kresin (Plenum, New York)}.\\ \\

\vspace*{2.cm}
\noindent
{\bf PAIR CORRELATIONS IN SUPERFLUID HELIUM 3}\\
\\
\\
\hspace*{1in}Dieter Vollhardt\\
\\
\hspace*{1in}Theoretische Physik III\\
\hspace*{1in}Elektronische Korrelationen und Magnetismus\\ 
\hspace*{1in}Universit"t Augsburg\\
\hspace*{1in}86135 Augsburg, Germany\\
\\
\\
\noindent
{\bf  ABSTRACT}\\\\
In 1996 Lee, Osheroff and Richardson received the Nobel Prize for their 1971 discovery of
superfluid helium 3 -- a discovery which opened the door to the most fascinating system known in
condensed matter physics. The superfluid phases of helium 3, originating from pair condensation of helium 3 atoms,
turned out to be the ideal test-system for many fundamental concepts of modern physics, such
as macroscopic quantum phenomena,
(gauge-)symmetries and their spontaneous breakdown,
topological defects, etc.\
Thereby they enriched condensed matter physics enormously and contributed significantly to our understanding
of various other physical systems, from heavy fermion and high-$T_c$ superconductors all the way to neutron
stars and the early universe. A pedagogical introduction is presented.
\\\\\\
\noindent
{\bf 1~~THE HELIUM LIQUIDS}
\\\\
There are two stable isotopes of the chemical element helium: helium 3 and helium 4, conventionally denoted by 
$^{3}$He and $^{4}$He, respectively. From a microscopic point of view, helium atoms are structureless, 
spherical particles interacting via a two-body potential that is well understood. The attractive part
of the potential, arising from weak van der Waals-type dipole (and higher multipole) forces, causes
helium gas to condense into a liquid state at temperatures of 3.2 K and 4.2 K for $^{3}$He and $^{4}$He,
respectively, at normal pressure. The pressure versus temperature phase diagrams of $^{3}$He and $^{4}$He
are shown in Figs. 1.1 and 1.2. When the temperature is decreased even further one finds that the helium
liquids, unlike all other liquids, do not solidify unless a pressure of around 30 bar is applied. This is the first 
remarkable indication of macroscopic quantum effects in these systems. The origin of this unusual
behaviour lies in the quantum-mechanical uncertainty principle, which requires that a quantum particle
can never be completely at rest at given position, but rather performs a zero-point motion about the
average position. The smaller the mass of the particle and the weaker the binding force, the stronger
these oscillations are. In most solids the zero-point motion is confined to a small volume of only a
fraction of the lattice-cell volume. In the case of helium, however, two features
combine to prevent the formation of a crystalline solid with a rigid lattice structure: (i) the strong 
zero-point motion arising from the small atomic mass (helium is the second-lightest element in the
periodic table); and (ii) the weakness of the attractive interaction due to the high symmetry of these
simple atoms.
It is this very property of helium -- of staying liquid -- that makes it such a valuable system for observing
quantum behaviour on a macroscopic scale.
Quantum effects are also responsible for the strikingly different behaviours of $^{4}$He and $^{3}$He at even
lower temperatures. Whereas $^{4}$He undergoes a second-order phase transition into a state later shown to
be superfluid, i.e. where the liquid is capable of flowing through narrow capillaries or tiny pores
without friction, no such transition is observed in liquid $^{3}$He in the same temperature range (see
Figs. 1.1 and 1.2). The properties of liquid $^{3}$He below 1 K are nevertheless found to be increasingly
different from those of a classical liquid. It is only at a temperature roughly one thousandth of
the transition temperature of $^{4}$He that $^{3}$He also becomes superfluid, and in fact forms ${\it several}$
superfluid phases, each of which has a much richer structure\footnote {A comprehensive
treatment of superfluidity in $^3$He with a very extensive reference list can be found in the book by Vollhardt and W\"olfle (1990).}
than that of superfluid $^4$He.

{\begin{figure}[t]
\def\epsfsize#1#2{14.5cm}\epsfbox{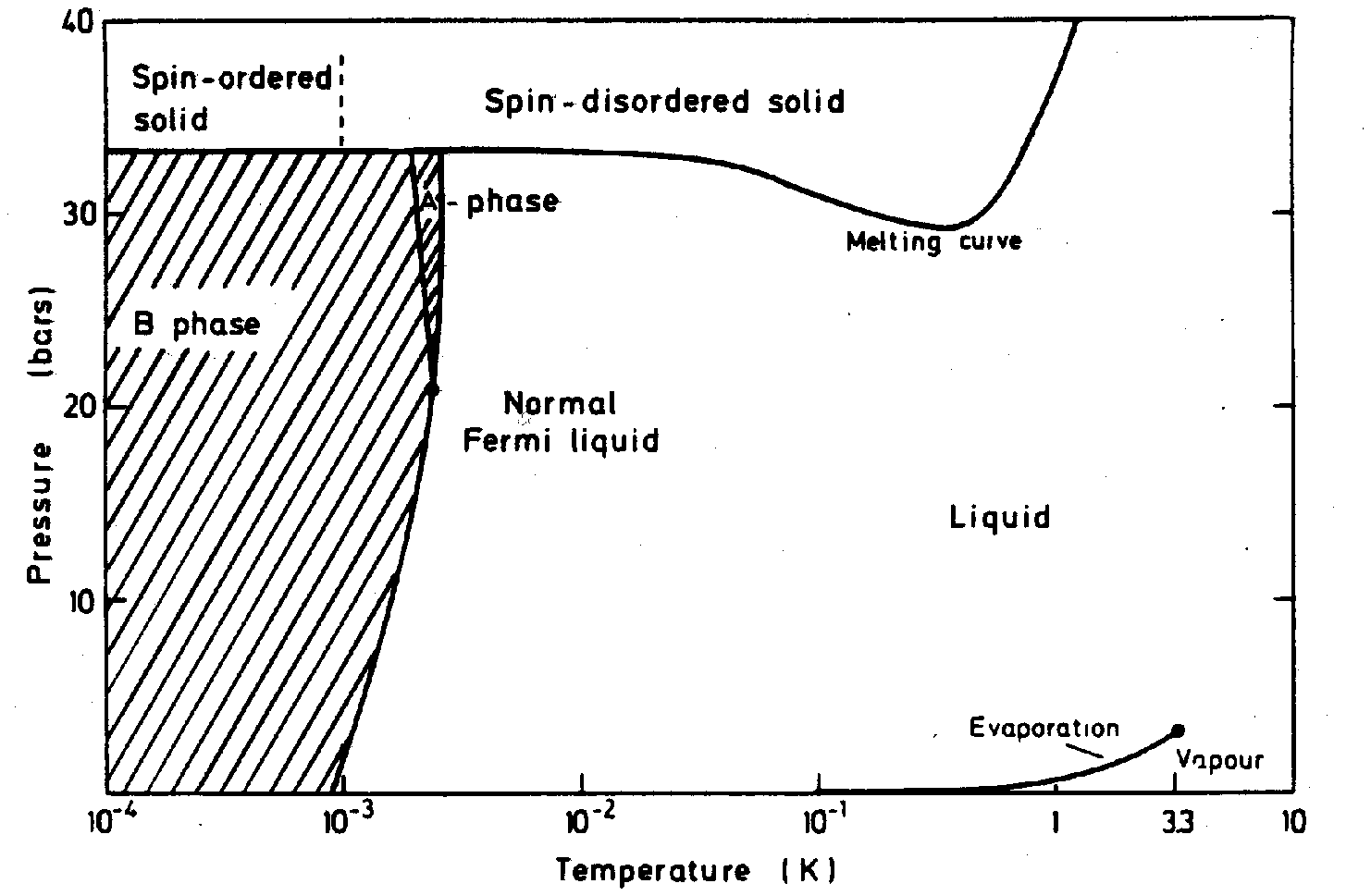}
\end{figure}
{\small\vspace*{-.5cm}
{\bf Figure 1.1}~~ Pressure versus temperature phase diagram for $^3$He; note the logarithmic temperature scale.}}
\vspace*{.5cm}

The striking difference in the behaviours of $^{3}$He and $^{4}$He at low temperatures is a consequence
of the laws of quantum theory as applied to systems of identical particles, i.e. the laws of quantum statistics. The
$^{4}$He atom, being composed of an even number of electrons and nucleons, has spin zero and consequently obeys Bose-Einstein 
statistics. In contrast, the $^{3}$He nucleus consists of $\em{three}$ nucleons, whose spins add up to give a total
nuclear spin of $I=\frac{1}{2}$, making the total spin of the entire $^{3}$He atom $\frac{1}{2}$ as well.
Consequently liquid $^{3}$He obeys Fermi-Dirac statistics. So it is the tiny nuclear spin, buried deep
inside the helium atom, that is responsible for all the differences of the macroscopic properties of
the two isotopes.

Since in a Bose system single-particle states may be multiply occupied, at low temperatures this system
has a tendency to condense into the lowest-energy single-particle state (Bose-Einstein condensation). It is
believed that the superfluid transition in $^{4}$He is a manifestation for Bose-Einstein condensation. The
all-important qualitative feature of the Bose condensate is its phase rigidity, i.e. the fact that it is 
energetically favourable for the particles to condense into a single-particle state of fixed quantum-mechanical
phase, such that the global gauge symmetry spontaneously broken. As a consequence, macroscopic flow of the
condensate is (meta-)stable, giving rise to the phenomenon of superfluidity.

In a Fermi system, on the other hand, the Pauli exclusion principle allows only single occupation of
fermion states. The ground state of the Fermi gas is therefore the one in which all single-particle states
are filled up to a limiting energy, the Fermi energy $E_{F}$. As predicted by Landau (1956, 1957, 1958) and later verified experimentally (for a review see Wheatley (1966)), the properties of $^{3}$He well 
below its Fermi temperature $\it{T}_{F} = \it{E}_{F}/\it{k}_{B}  {\approx}  1 K$ are similar to those
of a degenerate Fermi gas.
In particular the formation of a phase-rigid condensate is not possible in this 
framework. Until the mid-1950s a superfluid phase of liquid $^{3}$He was therefore believed to be ruled
out. On the other hand, it is most remarkable that the property of superfluidity (London, 1950, 1954) was
indeed first discovered experimentally in a ${\it Fermi}$ system, namely that of the ``liquid''  
of conduction electrons in a superconducting metal (Kamerlingh Onnes, 1911). The superfluidity of $^{4}$He was only found more than 25 years later.

{\begin{figure}[t]
\def\epsfsize#1#2{14.5cm}\epsfbox{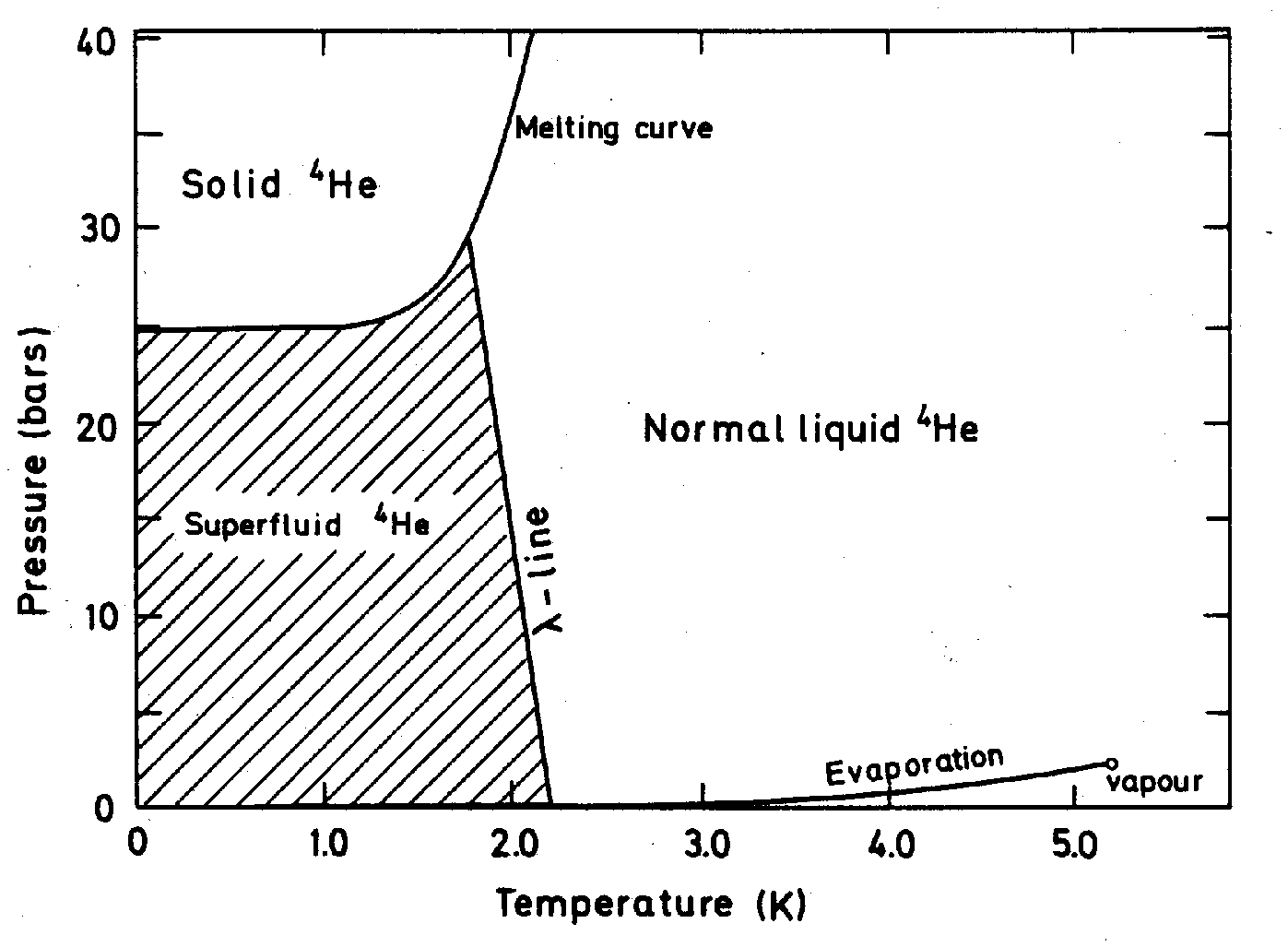}
\end{figure}
{\small\vspace*{-.5cm}
{\bf Figure 1.2}~~ Pressure versus temperature phase diagram for $^4$He; linear temperature scale.
}}
\vspace*{.5cm}
\\\\\\
\noindent
{\bf 2~~PAIR CONDENSATION IN A FERMI LIQUID}
\\\\
The key to the theory of superconductivity (Bardeen, Cooper and Schrieffer (BCS) 1957) turned out to be
the formation of ``Cooper pairs'', i.e. pairs of electrons with opposite momentum ${\bf k}$
and spin projection $\sigma$: (${\bf k}\uparrow, -{\bf k}\downarrow$). In the case of conventional
superconductors the Cooper pairs
are structureless objects, i.e. the two partners form a spin-singlet state in a relative s-wave
orbital state. These Cooper pairs have total spin zero and may therefore be looked upon in a way as composite bosons, which all have
the same pair wave function and are all in the same quantum-mechanical state. In this picture
the transition to the superconducting state corresponds to the Bose-condensation of Cooper pairs, the condensate being characterized by macroscopic quantum coherence.
The concept of Bose-Einstein condensation is appealing since key features of superconductivity like
the Meissner effect, flux quantization and superfluid mass currents in conventional superconductors
are naturally implied. Nevertheless, since the theory of conventional superconductivity is firmly
based on BCS theory, the concept of a Bose-Einstein condensation of Cooper pairs traditionally did not
receive much attention (or was even considered to be downright wrong). Within the context of superfluid
$^3$He, this notion was taken up again by Leggett (1980a,b), who argued that tightly bound Bose-Einstein-condensed molecules on the one hand and Cooper pairs on the other may be viewed as extreme limits of the
same phenomenon. This approach, which was quite provocative at the time, is now well accepted (Zwerger, 1992;
Nozi\`eres, 1995; Randeria, 1995). However, the original idea that at T$_c$ Cooper pairs form and automatically
Bose-condense has been revised since then. Apparently Cooper pair formation is not a separate phase
transition but is rather a matter of thermal equilibrium: for any finite coupling there exists a finite
density of pairs even above T$_c$, although in conventional superconductors -- and even in
high T$_c$ materials -- their density is negligibly small. At weak coupling (BCS limit) the condensation
temperature and the (not well-defined) temperature of pair formation practically coincide; they become
different only at very strong coupling (Bose limit). Similar ideas are also implicit in several
theoretical approaches to high-$T_c$ superconductivity.

While in free space an attractive force has to be sufficiently strong to bind two electrons, inside
the metal the presence of the filled Fermi sea of conduction electrons blocks the decay of a Cooper
pair, so that an ${\it arbitrarily}$ small attractive interaction leads to the formation of stable
Coopers pairs. The attractive interaction between the electrons of a Cooper pair in a conventional
superconducting metal is due to the exchange of virtual phonons (electron-phonon interaction). If the 
phonon-mediated interaction is strong enough to overcome the repulsive Coulomb interactions between
the two electrons then a transition into a superconducting state may occur. On the other hand, any
other mechanism leading to attraction between electrons at the Fermi surface is equally well suited
for producing superconductivity.

Given the success of the BCS theory in the case of superconductivity, it was natural to ask whether
a similar mechanism might also work for liquid $^{3}$He. Since there is no underlying crystal lattice
in the liquid that could mediate the attractive force, the attraction must clearly be an intrinsic
property of the one-component $^{3}$He liquid itself. The main feature of the interatomic $^{3}$He potential
is the strong repulsive component at short distances, and the weak van der Waals attraction at medium
and long distances. It soon became clear that, in order to avoid the hard repulsive core and thus make
optimal use of the attractive part of the potential, the $^{3}$He atoms would have to form Cooper-pairs
in a state of ${\it nonzero}$ relative angular momentum $l$. In this case the Cooper-pair wave function
vanishes at zero relative distance, thus cutting out the most strongly repulsive part of the potential.
In a complementary classical picture one might imagine the partners of a Cooper pair revolving about
their centre of gravity, thus being kept away from each other by the centrifugal force.

When the superfluid phases of $^{3}$He were finally discovered in 1971 at temperatures of about 2.6 mK and
1.8 mK  respectively (Osheroff, Richardson and Lee, 1972a), in an experiment actually designed to observe a magnetic
phase transition in solid $^{3}$He, the results came as a great surprise.
\\\\\\
\noindent
{\bf 3~~PROPERTIES OF SUPERFLUID $^{3}$He}
\\\\
Soon after the discovery of the phase transitions by Osheroff, Richardson and Lee (1972a), it was
possible to identify altogether ${\it three}$ distinct stable superfluid phases of bulk $^{3}$He
; these are referred to as the A,B and A$_{1}$ phases. In zero magnetic field only the
A and B phases are stable. In particular, in zero field the A phase only exists within a finite range
of temperatures, above a critical pressure of about 21 bar. Hence its region of stability in the
pressure-temperature phase diagram has a roughly triangular shape as shown in Fig. 1.1. The B phase, on the other hand, occupies the largest part of this phase diagram and is found to be stable down to
the lowest temperatures attained so far. Application of an external magnetic field has a strong 
influence on this phase diagram. First of all, the A phase is now stabilized down to zero pressure. 
Secondly, an entirely new phase, the A$_{1}$ phase, appears as a narrow wedge between the normal state
and the A and B phases.

Owing to the theoretical work on anisotropic superfluidity that had been carried out before the actual
discovery of superfluid $^{3}$He, progress in understanding the detailed nature of the phases was very
rapid. This was clearly also due to the excellent contact between experimentalists and theorists, which
greatly helped to develop the right ideas at the right time; for reviews see Leggett (1975), Wheatley (1975),
Lee and Richardson (1978). In particular, it fairly soon became
possible to identify the A phase and the B phase as realizations of the states studied previously by 
Anderson and Morel (1960, 1961) and Balian and Werthamer (1963) respectively. Therefore the A phase
is described by the so-called ``Anderson-Brinkman-Morel'' (ABM) state, while the B phase is described by the
``Balian-Werthamer'' (BW) state. Consequently, ``A phase'' and ``ABM state'' are now used as
synonyms; the same is true in the case of ``B phase'' and ``BW state''. (The fact that the ABM state
describes the A phase and the BW state the B phase is a very fortunate coincidence -- if it was the other
way around, it would be quite confusing!).

Although the three superfluid phases all have very different properties, they have one important thing
in common: the Cooper pairs in all three phases are in a state with ${\it parallel}$ spin (S = 1) and 
relative orbital angular momentum $l = 1$. This kind of pairing is referred to as ``spin-triplet p-wave
pairing''. In contrast, prior to the discovery of the superfluid phases of $^{3}$He, Cooper pairing
in superconductors was only known to occur in a state with opposite spins (S = 0) and $l = 1$, i.e. in a
``spin-singlet s-wave state''. It should be noted that Cooper pairs in a superconductor and in superfluid
$^{3}$He are therefore very different entities: in the former case pairs are formed by pointlike,
structureless electrons and are spherically symmetric, while in the case of $^{3}$He Cooper pairs are made of
actual atoms (or rather of quasiparticles involving $^{3}$He atoms) and have an internal structure themselves.
\\\\
\noindent
{\bf 3.1~~The internal structure of Cooper pairs}
\\\\
Quantum-mechanically, a spin-triplet configuration (S = 1) of two particles has three substates with
different spin projection $S_{z}$. They may be represented as $\mid \uparrow\uparrow \rangle$ with
$S_{z}= +1, 2^{-1/2}(\mid \uparrow \downarrow \rangle + \mid \downarrow \uparrow)$ with
$S_{z} = 0$,  and $\mid \downarrow\downarrow\rangle$ with $S_{z} = -1$. The pair wave function $\Psi$ is in general
a linear superposition of all three substates, i.e.
\begin{equation}
\Psi = \psi_{1,+}({\bf k}) \mid \uparrow\uparrow \rangle + \psi_{1,0}({\bf k})(\mid \uparrow \downarrow
\rangle + \mid \downarrow \uparrow \rangle) + \psi_{1,-}({\bf k}) \mid \downarrow\downarrow \rangle
\end{equation}
where $\psi_{1,+}({\bf k}), \psi_{1,0}({\bf k})$ and $\psi_{1,-}({\bf k})$
are the three complex-valued amplitudes of the respective substates. In the case of a superconductor,
where S = 0 and $l = 0$, the pair wave function is much simpler, i.e. it is given by only a single
component
\begin{equation}
\Psi_{sc} = \psi_{0}(\mid\uparrow\downarrow\rangle - \mid\downarrow\uparrow\rangle)
\end{equation}
with a single amplitude $\psi_{0}$.

So far we have only taken into account that, since S = 1, there are three substates for the spin. The 
same is of course true for the relative orbital angular momentum $l = 1$ of the Cooper pair, which also
has three substates $l_{z} = 0, \pm1$. This fact is important if we want to investigate the amplitudes
$\psi_{1,+}({\bf k})$ etc. further. They still contain the complete information about the space (or
momentum) dependence of $\Psi$. The pair wave function $\Psi$ is therefore characterized by three spin
substates and three orbital substates, i.e. by altogether 3 x 3 = 9 substates with respect to the spin
and orbital dependence. Each of these nine substates is connected with a complex-valued parameter. Here we
see the essential difference between Cooper pairs with $S = l = 0$ (conventional superconductors) and
$S = l = 1 (^{3}He)$: their pair wave functions are very different. In the former case a single
complex-valued parameter is sufficient for its specification, in the latter case of superfluid 
$^{3}$He  {\it nine} such parameters are required. This also expresses the fact that a Cooper pair in
superfluid $^{3}$He has an internal structure, while that for a conventional superconductor does not:
because $l = 1$, it is intrinsically {\it anisotropic}. This anisotropy may conveniently be described
by specifying some direction with respect to a quantization axis both for the spin and the orbital
component of the wave function.

In order to understand the novel properties of superfluid $^{3}$He, it is therefore important to keep
in mind that there are two characteristic directions that specify a Cooper pair. Here lies the substantial
difference from a superconductor and the origin of the multitude of unusual phenomena occurring in 
superfluid $^{3}$He: the structure of the Cooper pair is characterized by {\it internal degrees of
freedom}. Nevertheless, in both cases the superfluid/superconducting state can be viewed as the condensation 
of a macroscopic number of these Cooper pairs into the same quantum-mechanical state, similar to a
Bose-Einstein condensation, as discussed above.
\\\\
\noindent
{\bf 3.2~~Broken symmetry and the order parameter}
\\\\
In the normal liquid state Cooper pairs do not exist. Obviously in the superfluid a new state of order
appears, which spontaneously sets in at the critical temperature $T_{c}$. This particular transition
from the normal fluid to the superfluid, i.e. into the ordered state, is called ``continuous'', since the
condensate -- and hence the state of order -- builds up continuously. This fact may be expressed 
quantitatively by introducing an ``order-parameter'' that is finite for $T < T_{c}$ and zero for
$T \geq T_{c}$. A well-known example of such a transition is that from a paramagnetic to a ferromagnetic
state of a metal when the system is cooled below the Curie temperature. In the paramagnetic regime
the spins of the particles are disordered such that the average magnetization $\langle {\bf M}\rangle$
of the system is zero. By contrast, in the ferromagnetic phase the spins are more or less aligned and
$\langle {\bf M} \rangle$ is thus finite. In this case the system exhibits long-range order of the spins.
The degree of ordering is quantified by $| \langle {\bf M} \rangle |$, the magnitude of the magnetization.
Hence $| \langle {\bf M} \rangle |$ is called the ``order parameter'' of the ferromagnetic state. Clearly, the existence
of a preferred direction {\bf M} of the spins implies that the symmetry of the ferromagnet under spin
rotations is reduced (``broken'') when compared with the paramagnet: the directions of the spins are no
longer isotropically distributed, and the system will therefore no longer be invariant under a spin
rotation. This phenomenon is called ``spontaneously broken symmetry''; it is of fundamental importance in
the theory of phase transitions. It describes the property of a macroscopic system (i.e. a system in the
thermodynamic limit) that is in a state that does not have the full symmetry of the microscopic dynamics.

The concept of spontaneously broken symmetry also applies to superconductivity and superfluid $^{3}$He. 
In this case the order parameter measures the existence of Cooper pairs and is given by the probability 
amplitude for a pair to exist at a given temperature. It follows from the discussion of the possible
structure of a Cooper pair in superfluid $^{3}$He that the associated order parameter will reflect
this structure and the allowed internal degrees of freedom. What then are the spontaneously broken
symmetries in superfluid $^{3}$He?

As already mentioned, the interparticle forces between the $^{3}$He atoms are rotationally invariant
in spin and orbital space and, of course, conserve particle  number. The latter symmetry gives rise to
a somewhat abstract symmetry called ``gauge symmetry''. Nevertheless, gauge symmetry is spontaneously
broken in any superfluid or superconductor. In addition, in an odd-parity pairing superfluid, as in the case
of $^{3}$He, where $l = 1$, the pairs are necessarily in a spin-triplet state, implying that rotational
symmetry in spin space is broken, just as in a magnet. At the same time, the anisotropy of the Cooper-pair
wave function in orbital space calls for a spontaneous breakdown of orbital rotation symmetry, as
in liquid crystals. All three symmetries are therefore simultaneously broken in superfluid $^{3}$He.
This implies that the A phase, for example, may be considered as a ``superfluid nematic liquid crystal
with (anti)ferromagnetic character''. One might think that a study of the above mentioned broken
symmetries could be performed much more easily by investigating them separately, i.e. within the isotropic
superfluid, the magnet, the liquid crystal etc. itself. However, the combination of several {\it 
simultaneously} broken continuous symmetries is more than just the simple sum of the properties of all
these known systems. Some of the symmetries broken in superfluid $^{3}$He are ``relative'' symmetries,
such as spin-orbit rotation symmetry or gauge-orbit symmetry (Leggett, 1972,1973b; Liu and Cross, 1978).
Because of this, a rigid connection is established between the corresponding degrees of freedom of the
condensate, leading to long-range order only in the combined (and not in the individual) degrees of
freedom. This particular kind of broken symmetry, for  example the so-called ``spontaneously broken
spin-orbit symmetry'', gives rise to very unusual behaviour, as will be discussed later.

It is clear that in principle the internal degrees of freedom of a spin-triplet p-wave state allow
for many different Cooper-pair states and hence superfluid phases. (This is again different from 
ordinary superconductivity with $S = 0, l = 0$ pairing, where only a {\it single} phase is possible).
Of these different states, the one with the lowest energy for given external parameters will be realized. 
In fact, Balian and Werthamer (1963) showed, that, within a conventional ``weak-coupling'' approach,
of all possible states there is precisely one state (the ``BW state'') that has the lowest energy at {\it all}
temperatures. This state is the one that describes the B phase of superfluid $^{3}$He. The state
originally discussed by these authors is one in which the orbital angular momentum ${\it\bf l}$ and spin
${\it\bf S}$ of a Cooper pair couple to a total angular momentum ${\it\bf J} = {\it\bf l} + {\it\bf S} = 0$. This
$^{3}P_{0}$ state is, however, only a special case of a more general one with the same energy (in the
absence of spin-orbit interaction), obtained by an arbitrary rotation of the spin axes relative to the
orbital axes of the Cooper-pair wave function. Such a rotation may be described mathematically by 
specifying a rotation axis ${\hat{\bf n}}$ and a rotation angle ${\theta}$. In the BW state all three
spin substates in (1) occur with equal measure. This state has a rather surprising property: in spite
of the intrinsic anisotropy, the state has an ${\it isotropic}$ energy gap. (The energy gap is the amount
by which the system lowers its energy in the condensation process, i.e. it is the minimum energy required
for the excitation of a single particle out of the condensate.) Therefore the BW state resembles ordinary
superconductors in several ways. On the other hand, even though the energy gap is isotropic, the BW state
is intrinsically anisotropic. This is clearly seen in dynamic experiments in which the Cooper-pair structure
is distorted. For this reason the BW state is sometimes referred to as ``pseudo-isotropic''. Owing to
the quantum coherence of the superfluid state, the rotation axis ${\hat{\bf n}}$ and angle ${\theta}$
characterizing a Cooper pair in the BW state are macroscopically defined degrees of freedom, whose
variation is physically measurable.

Since in weak-coupling theory the BW state always has the lowest energy, an explanation of the
existence of the A phase of superfluid $^{3}$He obviously requires one to go beyond such an approach
and to include ``strong--coupling effects''(Anderson and Brinkmann, 1973, 1978; for a review of a
systematic approach see Serene and Rainer (1983)). In view of the fact that at present microscopic theories 
are not capable of computing transition temperatures for $^{3}$He, it is helpful to single out a particular
effect that can explain the stabilization of the A phase over the B phase at least qualitatively. As shown by Anderson and Brinkman (1973), there is such a conceptually simple effect, which is based on a feedback mechanism: the
pair correlations in the condensed state change the pairing interaction between the $^{3}$He quasiparticles,
the modification depending on the actual state itself. As a specific mechanism, these authors considered
the role of spin fluctuations and showed that a stabilization of the state first considered by
Anderson and Morel (1960, 1961) is indeed possible. This only happens at somewhat elevated pressures, since spin
fluctuations become more pronounced only at higher pressures. This ``ABM state'' (from the initials of the above three authors) does indeed describe the A phase. It has
the property that, in contrast with $^{3}$He-B, its magnetic susceptibility is essentially the same
as that of the normal liquid. This is a clear indication that in this phase the spin substate with
$S_{z}$ = 0, which is the only one that can be reduced appreciably by an external magnetic field, is
absent. Therefore $^{3}$He-A is composed only of $\mid\uparrow\uparrow\rangle$ and $\mid\downarrow
\downarrow\rangle$ Cooper pairs. This implies that the anisotropy axis of the spin part of the Cooper
pair wave function, called $\hat{\bf d}$, has the same fixed direction in every pair. (More precisely,
$\hat{\bf d}$ is the direction along which the total spin of the Cooper pair vanishes:
$\hat{\bf d} \cdot \bf {S}$ = 0). Likewise, the direction of the relative orbital angular momentum 
$\hat {\bf l}$ is the same for all Cooper pairs. Therefore in the A phase the anisotropy axes
$\hat{\bf d}$ and $\hat{\bf l}$ of the Cooper-pair wave function are long-range-ordered, i.e. are
preferred directions in the whole macroscopic sample. This implies a pronounced anisotropy of this
phase in all its properties. In particular, the value of the energy gap now explicitly depends on the
direction in $\bf{k}$ space on the Fermi sphere and takes the form
\begin{equation}
\Delta_{\hat{\bf k}}(T) = \Delta_{0}(T)[1 - (\hat{\bf k} \cdot \hat{\bf l})^{2}]^{1/2}.
\end{equation} 
Hence the gap vanishes at two points on the Fermi sphere, namely along $\pm \hat{\bf l}$. Because of the
existence of an axis $\hat{\bf l}$, this state is also called the ``axial state''. The existence of
nodes implies that in general quasiparticle excitations may take place at arbitrarily low temperatures.
Therefore, in contrast with $^{3}$He-B or ordinary superconductors, there is a finite density of states
for excitations with energies below the average gap energy, leading for example to a specific heat
proportional to $T^{3}$ at low temperatures.

The third experimentally observable superfluid phase of $^{3}$He, the $A_1$ phase, is only stable in the
presence of an external magnetic field. In this phase Cooper pairs are all in a single spin substate, the
$\mid \uparrow\uparrow\rangle$ state, corresponding to $S_z$ = + 1; the components with $\mid\uparrow
\downarrow\rangle$ + $\mid\downarrow\uparrow \rangle$ and $\mid\downarrow\downarrow\rangle$ states are
missing. It is therefore a ``magnetic'' superfluid, the first ever observed in nature.
\\\\
\noindent
{\bf 3.3~~Orientational effects}
\\\\
For a pair-correlated superfluid, the pairing interaction is the most important interaction, since it
is responsible for the formation of the condensate itself. Nevertheless, there also exist other, much
weaker, interactions, which may not be important for the actual transition to the pair-condensed state,
but which do become important if their symmetry differs form the aforementioned. In particular, they
may be able to break remaining degeneracies.
\\\\
\noindent
{\bf The dipole--dipole interaction}.
The dipole--dipole interaction between the nuclear spins of the $^{3}$He atoms leads to a very weak,
spatially strongly anisotropic, coupling. The relevant coupling constant $g_D(T)$ is given by
\begin{equation}
g_D(T) \approx \frac{\mu^2_0}{a^3} \left(\frac{\Delta(T)}{E_F}\right)^{2} n
\end{equation} 
Here $\mu_0$ is the nuclear magnetic moment, such that $\mu^2_0/a^3$ is the average dipole energy of
two particles at relative distance $a$ (the average atomic distance), while the second factor measures the
probability for these two particles to form a Cooper pair and $n$ is the overall particle density.
Since $\mu^2_0/a^3$ corresponds to about  $10^{-7}$K, this energy is extremely small and the resulting
interaction of quasiparticles at temperatures of the order of $10^{-3}$K might be expected to be
completely swamped by thermal fluctuations. This is indeed true in a normal system.
However, the dipole-dipole interaction implies a spin-orbit coupling and thereby has a symmetry different
from that of the pairing interaction. In the condensate the symmetries with respect to a rotation in spin
and orbital space are spontaneously broken, leading to long-range order (for example of $\bf\hat d$ and
$\bf\hat l$ in the case of $^3$He-A). Nevertheless, the pairing interaction does not fix the 
$relative$ orientation of these preferred directions, leaving a continuous degeneracy. As pointed out
by Leggett (1973a,b, 1974, 1975), in this situation the tiny dipole interaction is able to lift the
degeneracy, namely by choosing that particular relative orientation of the long-range ordered preferred
directions for which the dipolar energy is
minimal. Thereby this interaction becomes of $macroscopic$ importance. One may also view this effect as
a $permanent$ local magnetic field of about 3 mT at any point in the superfluid (in a liquid!). In 
$^3$He-A the dipolar interaction is minimized by a parallel orientation of $\bf\hat d$ and $\bf\hat l$.
\\\\
\noindent
{\bf Effect of a magnetic field}.
An external magnetic field acts on the nuclear spins and thereby leads to an orientation
of the preferred direction in spin space. In the case of $^3$He-A the orientation energy is minimal if
$\bf\hat d$ is perpendicular to the field $\bf H$, since (taking into account $\bf\hat d \cdot \bf S$
= 0) this orientation guarantees ${\bf S} {\Vert} {\bf H}$.
\\\\
\noindent
{\bf Walls}.
Every experiment is performed in a volume of finite size. Clearly, the walls will have some effect on 
the liquid inside. In superfluid $^3$He this effect may readily be understood by using a simple picture.
Let us view the Cooper pair as a kind of giant ``molecule'' of two $^3$He quasiparticles orbiting around
each other. For a pair not to bump into a wall, this rotation will have to take place in a plane parallel
to the wall. In the case of $^3$He-A, where the orbital angular momentum $\bf\hat l$ has the same
direction in all Cooper pairs (standing perpendicular on the plane of rotation), this means that
$\bf\hat l$ has to be oriented perpendicular to the wall. So there exists a strict orientation of
$\bf\hat l$ caused by the walls (Ambegaokar et al., 1974). In the B phase, with its (pseudo)
isotropic order parameter, the orientational effect is not as pronounced, but there are qualitatively
similar boundary conditions. 
\\\\
\noindent
{\bf 3.4~~Textures}
\\\\
From the above discussion, it is clear that the preferred directions $\bf\hat l$ and $\bf\hat d$ in
$^3$He-A are in general subject to different, often competing, orientational effects (for simplicity,
we shall limit our description to $^3$He-A). At the same time, the condensate will oppose any spatial
variation of its long-range order. Any ``bending'' of the order-parameter field will therefore increase
the energy, thus giving an internal stiffness or rigidity to the system. While the orientational effects
might want $\bf\hat d$ and $\bf\hat l$ to adjust on the smallest possible lengthscale, the bending
energy wants to keep the configuration as uniform as possible. Altogether, the competition between these
two opposing effects will lead to a smooth spatial variation of $\bf\hat d$ and $\bf\hat l$ throughout 
the sample, called a ``texture''. This nomenclature is borrowed from the physics of liquid crystals,
where similar orientational effects of the preferred directions occur.

The bending energy and all quantitatively important orientational energies are invariant under the
replacement $\bf\hat d\rightarrow -\bf\hat d, \bf\hat l\rightarrow -\bf\hat l$. A state where
$\bf\hat d$ and $\bf\hat l$ are parallel therefore has the same energy as one where $\bf\hat d$ and $\bf\hat l$ are 
antiparallel. This leads to two different, degenerate, ground states. There is then the possibility
that in one part of the sample the system is in one ground state and in the other in a different
ground state. Where the two configurations meet they form a planar ``defect'' in the texture, called
a ``domain wall'' (Maki, 1977). This is in close analogy to the situation in a ferromagnet composed
of domains with different orientations of the magnetization. Domain walls are spatially localized and 
are quite stable against external perturbations. In fact, their stability is guaranteed by the specific nature 
of the order-parameter structure of $^3$He-A. Mathematically, this structure may be analysed according to
its topological properties; for reviews see Mermin (1979) and Mineev (1980). The stability of a domain wall can then be traced back to the existence of
a conserved ``topological charge''. Using the same mathematical approach, one can show that 
the order-parameter fields of the superfluid phases of $^3$He not only allow for planar defects but also for point
and line defects, called ``monopoles'' and ``vortices'' respectively. Defects can be ``nonsingular''
or ``singular'' , depending on whether the core of the defect remains superfluid or whether it is
forced to become normal liquid. The concept of vortices is of course well known from superfluid
$^4$He. However, since the order-parameter structure of superfluid $^3$He is so much richer than that of
superfluid $^4$He, there exist a wide variety of different vortices in these phases. Their detailed
structure has been the subject of intensive investigation, in particular in the context of experiments
on rotating superfluid $^3$He, where they play a central role (Hakonen and Lounasmaa, 1987; Salomaa and
Volovik, 1987).
\\\\
\noindent
{\bf 3.5~~Dynamic properties}
\\\\
From the discussion presented so far, we have already seen that the static properties of an anisotropic
superfluid are very unusual. Clearly, the dynamic properties can be expected to be at least as new and
diverse. Indeed, the fact that in superfluid $^3$He Cooper pairs have an internal structure can only be
investigated in detail by studying the dynamics, i.e. the frequency and momentum dependence, of the 
condensate. One may roughly distinguish between magnetic and nonmagnetic dynamic properties, depending
on whether the magnetization of the system is probed or whether properties such as mass transport or
the propagation of sound are studied.

For the investigation of dynamical effects it is instructive to have an idea of the typical frequencies
inherent to the superfluid condensate. Both the normal and the superfluid components are essentially
characterized by a single timescale each: for the normal component this is the quasiparticle lifetime
$\tau$, and for the superfluid component it is $\hbar/\Delta(T)$, where $\Delta(T)$ is the average
of the temperature-dependent energy gap. The orders of magnitude of the equivalent frequencies are
given by $\tau^{-1} \approx$ 10 MHz and $\Delta(T)/\hbar \approx 10^3 (1 - T/T_c)^{1/2}$ MHz,  
i.e. usually one has $\tau^{-1} \ll \Delta(T)/\hbar$. For frequencies $\omega$ much smaller than either of
these characteristic values, the liquid is always in local thermodynamic equilibrium, since the system 
always has sufficient time to adjust to any change induced on the timescale $\omega^{-1}$. This is
called the ``hydrodynamic regime'', which is important for a couple of reasons: (i) in this regime
knowledge of the conserved quantities and of those describing the broken symmetries is sufficient
to describe the properties of the system; and (ii) this regime is experimentally well
accessible. The multitude of broken symmetries in superfluid $^3$He consequently leads to very rich 
hydrodynamics, which describes the various low-frequency collective excitations of the system. Here
the word ``collective'' (as opposed to ``single-particle'') means that a macroscopic number of particles
is involved in a coherent fashion.
\\\\
\noindent
{\bf Spin dynamics}.
Investigations of the collective magnetic (i.e. spin-dependent) properties of the superfluid phases
of $^3$He by nuclear magnetic resonance (NMR) were particularly useful in identifying the explicit
order-parameter structure of these phases (Lee and Richardson, 1978). In usual NMR experiments the system under investigation
is brought into a strong constant external magnetic field ${\bf H}_0 = H_0 \hat {\bf z}$, which forces
the (nuclear) spin $\bf S$ to precess about ${\bf H}_0$. By applying a weak high-frequency magnetic
field ${\bf H}_{rf}$ perpendicular to ${\bf H}_0$, one is able to induce transitions in $S_z$, the
component along ${\bf H}_0$, of magnitude ${\pm\hbar}$. This effect is observed as an energy absorption
from the magnetic field. In the case of noninteracting spins these transitions occur {\it exactly} at the
energy ${\gamma\hbar H_0}$, i.e. at the Larmor frequency $\omega_L = \gamma H_0$, where $\gamma$ is the
gyromagnetic ratio of the nucleus. How does this change in the presence of interactions? For a spin
of magnitude $\frac {1}{2}\hbar$, as in the case of the $^3$He nucleus, a very general statement is 
possible (Leggett, 1972, 1973b): as long as the interactions are spin-{\it conserving}, there is no
change at all -- the resonance remains at $\omega_0$. On the other hand, for spin-nonconserving interactions, such as the spin-orbit interaction caused by the dipole coupling of the nuclear spins, a frequency shift
may indeed occur. However, such a ``nonsecular'' shift will usually be very small, namely at most of the    
order of the linewidth. The experimental data obtained by Osheroff et al. (1972b) in connection
with their discovery of the superfluid phases therefore came as a great surprise - they found that the
resonance, although still very sharp, occurred at frequencies substantially higher than $\omega_L$. The 
origin of this large shift was especially mysterious, since it obviously corresponded to a {\it constant}
local magnetic field of order 3 mT surrounding the nuclear spins in the liquid.

The solution to this puzzle was found by Leggett (1972, 1973b, 1974, 1975), who showed that the NMR shifts
are a consequence of the broken symmetries of the spin-triplet p-wave condensate, which he named
``spontaneously broken spin-orbit symmetry''. As explained earlier, the meaning of this concept
is that the preferred directions in spin and orbital space are long-range-ordered (individually so, or
in a combined way) and the tiny dipole interaction may take advantage of this situation by lifting the 
remaining degeneracy. The macroscopic quantum coherence of the condensate therefore raises the dipole
coupling to macroscopic importance. In this way, Leggett (1974) was able to calculate the general NMR
response of the spin-triplet p-wave condensate. In particular, in the A phase the transverse NMR frequency
$\omega_t$ is given by
\begin{equation} \omega^2_t = \omega^2_L + \Omega^2_A(T)\nonumber
\end{equation}
where $\Omega^2_A(T)$ is proportional to the dipole coupling constant (see (4)). It should be noted
that the field and temperature dependences of $\omega_t$ are neatly separated in a ``Pythagorean'' form:
$\omega_L$ only depends on $H_0$ and $\Omega_A$ only on $T$. In fact, Leggett (1974) worked out a complete
theory of spin dynamics, whose predictions were experimentally confirmed in every detail. For example, the
equation of motion of the total spin ${\bf S}$  is given by
\begin{equation}
\dot{\bf S} = \gamma{\bf S} \times {\bf H} + {\bf R}_D,
\end{equation}
where ${\bf H} = {\bf H}_0 + {\bf H}_{rf}$ is the total external magnetic field and a dot over a symbol
indicates the time derivative. Here ${\bf R}_D$ is the anisotropic ``dipole torque'', which itself
depends on the change of the dipole energy under a reorientation of the order parameter. In the normal
phase ${\bf R}_D$ is always zero. In the superfluid, however, one has ${\bf R}_D \not=0$, except for static
situations. If the system is displaced from static equilibrium (for example by applying ${\bf H}_{rf}$), ${\bf R}_D$ 
acts as restoring force. For example, in the A phase a periodic oscillation of ${\bf S}$ will lead
to an oscillation of $\hat{\bf d}$, the preferred direction in spin space, around the orbital degree of
freedom $\hat{\bf l}$ (which may be assumed to remain fixed because it cannot move very quickly). 
Equation (6) led Leggett to a spectacular prediction: even if the high-frequency field ${\bf H}_{rf}$
is oriented {\it parallel} to ${\bf H}_0$, there is a resonance, i.e. there exists a {\it longitudinal} spin
resonance! Since in this case $({\bf S} \times {\bf H})_z = 0$, (6) yields $dS_z/d{\it t} = {\it R}_{Dz}$.
In a normal system there can be no resonance since there is no restoring force: the z-component of the 
magnetization will simply relax exponentially but will not oscillate. How then can we understand the nature
of the longitudinal oscillation in the case of superfluid $^3$He-A? The A phase only consists of the
two spin substates $\vert\uparrow\uparrow\rangle$ and $\vert\downarrow\downarrow\rangle$. They may be
viewed as essentially independent interpenetrating superfluids, which are only very weakly coupled by
the spin-nonconserving dipole coupling. This coupling allows for a transition of $\vert\uparrow\uparrow
\rangle$ pairs into $\vert\downarrow\downarrow\rangle$ pairs, and vice versa. (The situation is quite
similar to a pair of weakly coupled superconductors, where Cooper pairs can tunnel from one superconductor to
the other (the ``Josephson effect''); the difference here is that the two subsystems fill the same 
volume, i.e. they are not spatially separated). Applying a high-frequency magnetic field parallel to the
static field ${\bf H}_0$ leads to oscillatory nonequilibrium between the two spin subsystems, with the
dipole interaction acting as a restoring force. The resonant frequency of this longitudinal oscillation occurs
at
\begin{equation}
\omega_l = \Omega_A(T),
\end{equation}
where $\Omega_A$ is the frequency that has already appeared in the expression for the transverse 
frequency (5).

Any texture formed by the order-parameter field changes the dipole torque ${\bf R}_D$ in a very specific
way. Therefore the measurement of NMR shifts, in combination with the corresponding theory, provides the
most versatile, and at the same time sensitive, tool for the investigation of order-parameter textures.

NMR frequencies are generally considerably smaller than the characteristic frequencies $\tau^{-1}$ and
$\Delta(T)/\hbar$ of the normal and superfluid components. Hence such experiments take place in the
hydrodynamic regime. At such low frequencies, i.e. energies, the magnitude of the order parameter
$\Delta(T)$ does not change at all - only the orientation of its spin part varies. Hence the {\it structure}
of the order  parameter is left intact - the dynamics is due to a ``rigid'' excitation of the order
parameter. At higher frequencies, $\omega \approx \Delta(T)$, this changes dramatically. To understand
the consequences of this, it is again helpful to view a Cooper pair as some kind of diatomic molecule.
As in the case of a molecule, an energy of the order of the binding energy will lead to internal 
excitations such as rotational and vibrational states.
\\\\
\noindent
{\bf Ultrasound excitations}.
Such a situation occurs in experiments measuring the attenuation of ultrasound at sound frequencies
close to $\Delta(T)/\hbar$ . Quite unexpectedly, one finds that the sound attenuation of the superfluid
has a sharp maximum directly below the transition temperature $T_c$; this maximum depends strongly on the
frequency $\omega$. These and other phenomena are explained by collective excitations of the order-parameter structure of the condensate. They owe their existence to pair correlations in a state with
nonzero relative orbital angular momentum, which imply an internal structure of the Cooper pair
(W\"olfle, 1973, 1978). This structure allows for the excitation of high-frequency $(\omega \approx \Delta
(T)/\hbar)$ collective oscillations (pair-vibration modes). Besides this, there is also the possibility
of a break-up of the Cooper pair. Pair breaking is only possible if the energy $\hbar\omega$ of the
sound wave is larger than the minimum energy for breaking a pair, 2$\Delta_{\hat{\bf k}}(T)$. Here
$\Delta_{\hat{\bf k}}(T)$ is the energy gap, which in general depends on $\hat{\bf k}$, the position on the
Fermi sphere. For smaller energies, only vibrations can be excited. A detailed theory of sound
absorption, including damping effects etc., has been developed (for details see Vollhardt and W\"olfle 
(1990)) and is in good agreement with experiments (Halperin and Varoquaux, 1990).
In particular, the existence of isotropic and anisotropic energy gaps in the B and A phase, respectively,
led to an early identification of these phases. Indeed, in the B phase sound attenuation is independent of
the direction of the sound entering the probe. By contrast, in the A phase it strongly depends on the
relative orientation of the sound wave to the anisotropy axis $\hat{\bf l}$. This orientation dependence
is very remarkable: by coupling to the nuclear spins, a weak external magnetic field of the order of
3 mT is able to change the direction of $\hat{\bf l}$ and thereby to modify the sound absorption. It is
the coherent ordering of {\it nuclear} spins that is ultimately responsible for the anisotropy of sound
absorption!
\\\\\\
\noindent
{\bf 4~~RELATION TO OTHER FIELDS AND RECENT DEVELOPMENTS}
\\\\
Why spend so much effort on sorting out the strange behavior of states of matter that
are not even found in nature, at temperatures well outside the reach of even a well-equipped low-temperature laboratory? Partly, of course, ``because it's there'', and because -- like any other system --
superfluid $^3$He deserves to be studied in its own right. However, what is even more important is that
superfluid $^3$He is a model system that exemplifies many of the concepts of modern theoretical
physics and, as such, has given us, and will further provide us, with new insights into the functioning of
quantum-mechanical many-body systems close to their ground state.

As discussed in Section 3.2, the key to understanding superfluid $^3$He is ``spontaneously broken symmetry''.
In this respect there are also fundamental connections with particle physics, deriving from the 
interpretation of the order-parameter field as a quantum field with a rich group structure. The collective
modes of the order-parameter as well as the localized topological defects in a given ground-state
configuration are the particles of this quantum field theory. Various anomalies known from particle
physics can be identified in the $^3$He model system, and one may hope that insights gained from the
study of superfluid $^3$He will turn out to be useful in elementary particle theory (Volovik, 1987, 1992).

There are several other physical systems for which the ideas developed in the context of superfluid
$^3$He are relevant. 
One or them is an anisotropic superfluid system that already exists in nature but is not accessible for 
laboratory experiments: this is the  nuclear matter forming the cores of neutron stars. There the pairing
of neutrons has been calculated to be of p-wave symmetry. Because of the strong spin-orbit  nuclear force,
the total angular momentum of the Cooper pairs is $J$ = 2 (Sauls et al., 1982; Pines and Alpar, 1985).

Above all, an anisotropic superconducting state is particularly exciting. There are now strong indications
that superconductivity in the so-called ``heavy-fermion'' systems and in high-$T_c$ cuprates, is, at least in some cases, due to the formation of anisotropic pairs with d-wave
symmetry (Cox and Maple, 1995). Many of the concepts and ideas developed for superfluid $^3$He have been
adapted to these systems.

The above discussion shows that superfluid $^3$He is a field of continuing interest.
Indeed, most recently superfluid $^3$He has been used as a test system for
the creation of ``cosmic strings" in the early stages of the universe.
According to Kibble (1976) and Zurek (1985) the observed inhomogeneity
of matter in the universe may be understood as the result of the
creation of defects generated by a rapid cooling through second-order
phase transitions, which led to the present symmetry-broken state of the
universe. In two different experiments with superfluid $^3$He, performed at
the low-temperature laboratories in Grenoble (B"uerle et al., 1996)
and Helsinki (Ruutu et al., 1996), a nuclear
reaction in the superfluid, induced by neutron radiation, caused a local
heating of the liquid into the normal state. During the subsequent,
rapid cooling back into the superfluid state the creation of a vortex
tangle was observed. The experimentally determined density of this
defect state was found to be consistent with Zurek's estimate and thus
gives important support to this cosmological model. Furthermore, a
recent experimental verification of momentogenesis in $^3$He-A by Bevan et al. (1997) was found to
support current ideas on cosmological baryogenesis. (Baryogenesis during
phase transitions in the early universe is believed to be responsible
for the observed excess of matter over antimatter). In view of these
exciting new developments it may become possible in the future to model
and study cosmological problems in the low-temperature laboratory in much more detail.

Due to the intense experimental and theoretical research since 1971 the
superfluid phases of $^3$He now belong to the best-understood states of
matter (Vollhardt and W\"olfle, 1990). The unique richness of their structure continues to lead to new
aspects whose investigation provides unexpected insights.
\\\\\\
\noindent
{\bf ACKNOWLEDGMENT}
\\\\
I am grateful to Peter W\"olfle for many useful discussions.
\\
\\
\\
\noindent
{\bf REFERENCES}
\\
\vspace{-.5em}
\\
\parindent0.1em
Ambegaokar, V., de Gennes, P.G., and Rainer, D., 1974, {\it Phys. Rev.} {\bf A9}, 2676.

Anderson, P.W., and Brinkman, W.F., 1973, {\it Phys. Rev. Lett.} {\bf 30}, 1108.

Anderson, P.W., and Brinkman, W.F., 1978, in {\it The Physics of Liquid and Solid Helium,
Part II}, K.H. Bennemann\\
\hspace*{1cm}  and J.B. Ketterson, eds., Wiley, New York, p. 177.

Anderson, P.W., and Morel, P., 1960, {\it Physica} {\bf 26}, 671.

Anderson, P.W., and Morel, P., 1961, {\it Phys. Rev.} {\bf 123}, 1911.

Balian, R., and Werthamer, N.R., 1963, {\it Phys. Rev.} {\bf 131}, 1553.

Bardeen, J., Cooper, L.N., and Schrieffer, J.R., 1957, {\it Phys. Rev.} {\bf 108}, 1175.

B\"auerle, C., Bunkov, Yu.M., Fisher, S.N., Godfrin, H., and Pickett, G.R., 1996,
{\it Nature} {\bf 382}, 332.

Bevan, T.D.C., Manninen, A.J., Cook, J.B., Hook, J.R., Hall, H.E., Vachaspati,
T., and Volovik, G.E., 1997,
{\it Nature}\\
\hspace*{1cm}  {\bf 386}, 689.

Cox, D.L., and Maple, M.B., 1995, {\it Physics Today}, {\bf 48}, No. 2, 32.

Hakonen, P., and Lounasmaa, O.V., 1987, {\it Phys. Today} {\bf 40}, 70.

Halperin, W.P., and Varoquaux, E., 1990, in {\it Helium Three}, W.P. Halperin and L.P.
Pitaevskii, eds., North-Holland,\\
\hspace*{1cm}  Amsterdam, p. 353.

Kamerlingh Onnes, H., 1911, {\it Proc. R. Acad. Amsterdam} {\bf 11}, 168

Kibble, T.W.B., 1976, {\it J. Phys.}, A{\bf 9}, 1387.

Landau, L.D., 1956, {\it Zh. Eksp. Teor. Fiz.} {\bf 30}, 1058 [Sov. Phys. JETP {\bf 3}, 920
(1957)].

Landau, L.D., 1957, {\it Zh. Eksp. Teor. Fiz.}, {\bf 32}, 59 [Sov. Phys. JETP {\bf 5}, 101
 (1957)].

Landau, L.D., 1958, {\it Zh. Eksp. Teor. Fiz.}, {\bf 35}, 97 [Sov. Phys. JETP {\bf 8}, 70
(1959)].

Lee, D.M., and Richardson, R.C., 1978, in {\it The Physics of Liquid and Solid Helium,
Part II}, K.H. Bennemann and\\
\hspace*{1cm}  J.B. Ketterson, eds., Wiley, New York, p. 287.

Leggett, A.J., 1972, {\it Phys. Rev. Lett.} {\bf 29}, 1227.

Leggett, A.J., 1973a, {\it Phys. Rev. Lett.} {\bf 31}, 352.

Leggett, A.J., 1973b, {\it J. Phys.} {\bf C6}, 3187.

Leggett, A.J., 1974, {\it Ann. Phys. (N.Y.)} {\bf 85}, 11.

Leggett, A.J., 1975, {\it Rev. Mod. Phys.} {\bf 47}, 331.

Leggett, A.J., 1980a, in {\it Proceedings of the XVI Karpacz Winter School of
Theoretical Physics} (Lecture Notes in \\
\hspace*{1cm} Physics, Vol. 115), A. Pekalski and
J. Przyslawa, eds., Springer, Berlin, p. 13.

Leggett, A.J., 1980b, {\it J. Physique} {\bf 41}, Colloq. C-7, p. 19.

Liu, M., and Cross, M.C., 1978, {\it Phys. Rev. Lett.} {\bf 41}, 250.

London, F., 1950, {\it Superfluids, Vol. I}, Wiley, New York.

London, F., 1954, {\it Superfluids, Vol. II}, Wiley, New York.

Maki, K., 1977, {\it Physica} {\bf 90B}, 84.

Mermin, N.D., 1979, {\it Rev. Mod. Phys.} {\bf 51}, 591.

Mineev, V.P., 1980, in {\it Soviet Scientific Reviews, Sect. A., Physics Reviews, Vol. 2},
Harwood Academic Publishers,\\
\hspace*{1cm}  Chur, p. 173.

Nozi\`eres, P., 1995, in {\it Bose-Einstein Condensation}, A. Griffin et al., eds., Cambridge
University Press.

Osheroff, D.D., Richardson, R.C., and Lee, D.M., 1972a, {\it Phys. Rev. Lett.} {\bf 28},
 885.

Osheroff, D.D., Gully, W.J., Richardson, R.C., and Lee, D.M., 1972b, {\it Phys. Rev. Lett.} {\bf 29}, 920.

Pines, D., and Alpar, A., 1985, {\it Nature} {\bf 316}, 27.

Randeria, M., 1995, in {\it Bose-Einstein Condensation}, A. Griffin et al., eds., Cambridge University Press.

Ruutu, V.M.H., Eltsov, V.B., Gill, A.J., Kibble, T.W.B., Krusius, M., Makhlin, Yu. G., Pla\c{c}ais, B., Volovik, G.E., \\
\hspace*{1cm} and Xu, W., 1996, {\it Nature} {\bf 328}, 334.

Salomaa, M.M., and Volovik, G.E., 1987, {\it Rev. Mod. Phys.}, {\bf 59}, 533.

Sauls, J. A., Stein, D.L., and Serene, J.W., 1982, {\it Phys. Rev.} {\bf D25}, 976.

Serene, J.W., and Rainer, D., 1983, {\it Phys. Rep.} {\bf 101}, 221.

Vollhardt, D., and W\"olfle, P., 1990, {\it The Superfluid Phases of Helium 3}, Taylor and Francis, London.

Volovik, G.E., 1987, {\it J. Low Temp. Phys.} {\bf 67}, 301.

Volovik, G.E., 1992, {\it Exotic properties of superfluid $^3$He}, World Scientific, Singapore.

Wheatley, J.C., 1966, in {\it Quantum Fluids, Vol. VI}, D.F. Brewer, ed., North-Holland, Amsterdam, p. 183.

Wheatley, J.C., 1975, {\it Rev. Mod. Phys.} {\bf 47}, 415.

W\"olfle, P., 1973, {\it Phys. Rev. Lett.} {\bf 30}, 1169.

W\"olfle, P., 1978, in {\it Progress in Low Temperature Physics, Vol. VIIA},  D.F. Brewer, ed., North-Holland, Amsterdam,\\
\hspace*{1cm} p. 191.

Zurek, W.H., 1985, {\it Nature} {\bf 317}, 505.

Zwerger, W., 1992, {\it Ann. der Physik} {\bf 1}, 15.
\end{document}